\documentclass[letterpaper,11pt]{article} %DO NOT CHANGE THIS
\usepackage{fullpage}  %Required
\usepackage{amsmath,amssymb,cleveref,natbib}  %Required
\usepackage{helvet}  %Required
\usepackage{courier}  %Required
\usepackage{url}  %Required
\usepackage{graphicx}  %Required
\usepackage{enumitem,import}
\usepackage{booktabs}
\usepackage{multirow}
\usepackage[table,xcdraw,pdftex,dvipsnames]{xcolor}
\usepackage{subfigure}
\usepackage{algorithm,algcompatible,algpseudocode, algorithmicx,xspace}
\usepackage{wrapfig}
\usepackage{authblk}

\makeatletter
\algrenewcommand\ALG@beginalgorithmic{\footnotesize}
\makeatother

\algrenewcommand\algorithmicindent{0.9em}

\usepackage{xargs}                      % Use more than one optional parameter in a new commands
\newcommand{\comment}[1]{}

\newcommand{\getdis}{\textsc{GetDist}\xspace}
\newcommand{\getnbd}{\textsc{GetNbrs}\xspace}
\newcommand{\mkcom}{\textsc{MakeComplete}\xspace}
\newcommand{\getcnbd}{\textsc{GetCombinedNbrs}\xspace}
\newcommand{\getpd}{\textsc{PD}\xspace}
\newcommand{\getpddm}{\textsc{GetPDFromDisMat}\xspace}
\newcommand{\getdm}{\textsc{DistMat}\xspace}
\newcommand{\getbd}{\textsc{B-Dis}\xspace}
\newcommand{\getwd}{\textsc{W-Dis}\xspace}
\newcommand{\spd}{{shortest-path distance}\xspace}
\newcommand{\pd}{{persistence diagram}\xspace}
\newcommand{\pds}{{persistence diagrams}\xspace}
\newcommand{\nbd}{{neighborhood}\xspace}
\newcommand{\nbds}{{neighborhoods}\xspace}
\newcommand{\cnbd}{{combined-\nbd}\xspace}
\newcommand{\Cnbd}{{Combined-\nbd}\xspace}
\newcommand{\cnbds}{{combined-\nbds}\xspace}
\newcommand{\rbt}{{$\frac{r}{2}$}\xspace}
\newcommand{\ws}{{Wasserstein-$q$}\xspace}
\newcommand{\wst}{{Wasserstein-$2$}\xspace}
\newcommand{\bn}{{Bottleneck}\xspace}
\newcommand{\knu}[0]{\shortstack[c]{$k$-nbd\\of $u$}\xspace}
\newcommand{\knv}[0]{\shortstack[c]{$k$-nbd\\of $v$}\xspace}

\begin{document}
\title{Understanding and Predicting Links in Graphs: A Persistent Homology Perspective}

\author{Sumit Bhatia\thanks{IBM Research AI, New Delhi, India. sumitbhatia@in.ibm.com},  Bapi Chatterjee\thanks{Institute of Science and Technology, Austria. bhaskerchatterjee@gmail.com}, 
 Deepak Nathani\thanks{IIT Hyderabad, India. me15btech11009@iith.ac.in}  and Manohar Kaul\thanks{IIT Hyderabad, India. mkaul@iith.ac.in}}

\date{}
\maketitle

\begin{abstract}
Persistent Homology is a powerful tool in Topological Data Analysis (TDA) to capture topological properties of data succinctly at different spatial resolutions. For graphical data, shape, and structure of the neighborhood of individual data items (nodes) is an essential means of characterizing their properties. In this paper, we propose the use of persistent homology methods to capture structural and topological properties of graphs and use it to address the problem of link prediction. We evaluate our approach on seven different real-world datasets and offer directions for future work.
\end{abstract}

\section{Introduction}
%\subsection{Problem Description}
%\label{subsec:problemdef}
A graph data structure representing pairwise relations or interactions among individuals or entities recurs in diverse real-world applications such as social and professional networks, biological phenomena such as protein-protein interactions~\citep{moreno}, word co-occurrences~\citep{dc}, and citation and collaboration networks~\citep{citeseer}. In all these applications, understanding how the network evolves and being able to predict the formation of new, hitherto non-existent links is an important problem in the knowledge discovery pipeline and has crucial applications such as predicting target genes for cancer research~\citep{knit}, social network analysis, and recommendation systems.  

Consider a graph $G=(V,E)$, where $V = \{v_i\}^{n}_{i=1}$ is a finite set of nodes and $E = \{e_k\}^{m}_{k=1}$ is a finite set of edges. We denote the edge connecting the pair of node $v_i,v_j \in V$ by $e_{ij}$. We assume that in $G$ multiple edges for the same pair of nodes do not exist and there is no edge connecting a node to itself. If $G$ is a \textit{directed} graph, $e_{ij} \neq e_{ji}$, whereas, $e_{ij} = e_{ji}$ if $G$ is \textit{undirected}.

Let $U$ denote the set of all \textit{possible} edges in $G = (\{v_i\}^{n}_{i=1},\{e_k\}^{m}_{k=1})$. If $G$ is undirected, $|U| = C(n,2) = n(n-1)/2$, whereas, if $G$ is directed, $|U| = 2{\times}C(n,2) = n(n-1)$. The set $U-E$ is called the set of \emph{potential} links. Often, in real-world settings, only a small subset of links $u \in U$ will materialize in future with $|u| << |U|$. Given $G = (V;E)$, the task of identifying the edges $e \in u$ is challenging and requires understanding and modelling the differences between sets $u$ and $U-u$.  

\subsection{Existing Methods for Link-prediction}
\label{subsec:existmethod}
One of the simplest and the most frequently used methods for predicting a link between two nodes utilize the intuition that the likelihood of a link between two nodes is high if they share many common neighbors~\citep{adamicAdar,milneWitten}. These measures have empirically found to perform well for a variety of networks~\citep{kleinbergLinkPrediction} and are widely adopted due to their intuitive nature and ease of computation. However, such methods by definition, are limited to second order neighborhood of the source node and hence ignore the global structural information about the underlying network.  Different approaches that consider global information for link prediction include measures based on an ensemble of all paths (such as the Katz score~\citep{katz}), measures derived from conducting random walks over the graph~\citep{supervisedRandomWalk}, and variants of PageRank (e.g., SimRank~\citep{simRank}). Ensemble methods that complement the network structure information with external information such as text documents have recently proposed for context-sensitive dynamic link prediction~\citep{contextHT}. An alternate class of methods for link prediction that has recently gained prominence learns continuous vector representations of nodes in a graph such that nodes sharing similar structural properties are mapped close to each other in the latent space (e.g., LINE~\citep{line}, DeepWalk~\citep{deepwalk}, node2vec~\citep{node2vec}).

\subsection{Our Contributions}
\label{subsec:contribute}
We propose an alternative perspective for understanding the evolution of links in networks and predicting the formation of new links via \textit{persistent homology} -- an algebraic tool for measuring the features of shapes -- of a graph.

\textit{Homology} of a point set $X$ roughly characterizes it in terms of features like connected components, tunnels and voids that roughly correspond to the 0-dimensional, 1-dimensional and 2-dimensional \textit{homology groups}, respectively.  
When the nodes $v_i \in V$ are mapped to the points $x_i \in X$ for $1 \leq i \leq n$, homology of $X$ essentially exhibits $G$ in terms of its \textit{shape-features} like connected components, tunnels, voids, etc. formed by its nodes and edges. However, in practice, the properties of these features depend a lot on the resolution or scale at which they are being studied, and it is crucial to study them across a spectrum of spatial resolutions. {Persistent homology} comes across as an algebraic tool to serve exactly this purpose.   

Given two query nodes $v_i,v_j \in V$, we apply persistent homology to study the shape-features at different scales of the subgraphs of $G$ induced by the extended individual and \cnbds of $v_i$ and $v_j$. We contrast the shape-features of the subgraphs at varying resolutions. Given a graph $G = (\{v_i\}^{n}_{i=1},\{e_k\}^{m}_{k=1})$ and a pair of nodes $v_i,v_j \in V$, this procedure enables us to understand how the shape of $G$ changes with respect to an extended {\cnbd} of $v_i$ and $v_j$ due to the existence, or for that matter non-existence, of the edge $e_{ij}$.

Unlike commonly used heuristics for link prediction that utilize local structure (such as the measures proposed by Adamic and Adar~\citeyearpar{adamicAdar}; Milne and Witten~\citeyearpar{milneWitten}), studying the shape features of a Graph at varying resolutions helps us to capture the global structure information. Also, unlike other methods that explore the entire graph for the sake of employing global information~\citep{node2vec,deepwalk,simRank} our approach is adaptive: we study the combined neighborhood whose size varies depending on the sparsity of the graph. Thus, we also address the drawback of the large cost of exploration of the entire graph.  To summarize, our main contributions are the following: 
\begin{enumerate}
	\item First, we present the necessary mathematical notions	 to describe our method: the \textit{persistence diagram} of a graph (Section~\ref{sec:pd}).   
	\item Thereafter, we describe a novel approach to understand the existence of an edge in a graph by application of persistent homology. Essentially, we argue and explain that for a pair of nodes, the \pd of the subgraph induced by their extended \nbd should not change much by adding or removing a naturally existing edge (Section~\ref{sec:ouralgo}).
	\item We evaluate our proposed approach on seven different real-world network datasets from different domains and of varying sizes. We also present comparisons with four commonly used baseline methods for link prediction (Section~\ref{sec:exp}). Results of the experiments reveal that the proposed approach for link prediction achieves robust performance across all the datasets. Despite being relatively simple and computationally less expensive than other state-of-the-art methods, it achieves similar, if not better, performance for the link prediction task. We discuss the experimental results and comment on the limitations and possible future extensions of the presented work (Section~\ref{sec:discuss}).
	
\end{enumerate}

\subsection{A High Level Insight to our Method}\label{subsec:insight}

\begin{figure}[!t]	
	\centering\def\svgwidth{0.9\columnwidth}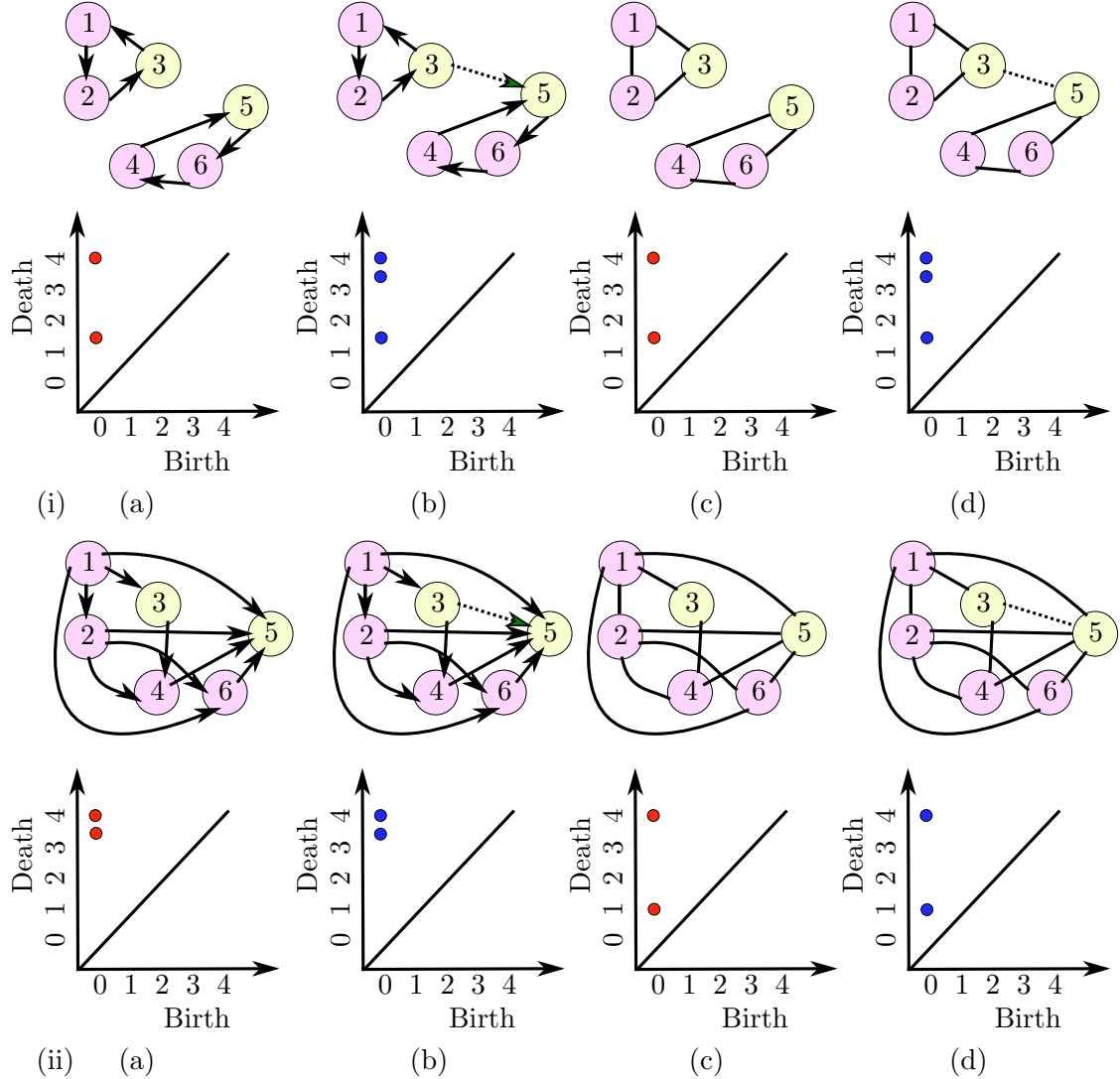
	\caption{Graphs and their persistence diagrams. Nodes with label 3 and 5 are query nodes.}
	\label{fig:toypd}
\end{figure}

Consider four toy examples of link prediction as shown in \ref{fig:toypd} where we illustrate how topological features capture changes in link structure of the graph. In all the figures, nodes 3 and 5 are the query nodes and we study how the persistence diagrams of the graph change when we induce an edge between these nodes (indicated by dashed edge). We consider four cases: \textit{(i)} a directed graph where the extended neighborhoods of query nodes (nodes 3 and 5) do not have any connecting link (\ref{fig:toypd}(i)(a,b)); \textit{(ii)} likewise for an undirected graph in \ref{fig:toypd}(i)(c,d); \textit{(iii)} a dense directed graph where the extended neighborhoods of query nodes have multiple links across (\ref{fig:toypd}(ii)(a,b)); and, \textit{(iv)} similarly for a dense undirected graph (\ref{fig:toypd}(ii)(a,b)).

The persistence diagrams of each of the graphs are placed directly below them in the \ref{fig:toypd}. The description of \pd is presented in the next section, here we only contrast them to provide intuition behind our method. We observe that in cases \textit{(i)} and \textit{(ii)}, addition of the induced edge causes significant changes in the \pd of the graph (note the blue and red points) indicating significant changes in the topology due to introduction of a link between two disconnected components of the graph.  Compare this with cases \textit{(iii)} and \textit{(iv)}, where the \pds remain unaffected by the addition of such a link. Intuitively, \textit{a link between two nodes is highly likely if there is a large overlap between their extended neighborhoods and addition of a link should not cause major changes in the topological properties of the graphs.} Motivated by this intuition, we wish to explore the changes in the \pds of extended \nbds of query nodes in order to predict a link between them and comparison of \pds of the induced subgraphs provides a robust tool for this purpose.

\subsection{Related Works on Application of Persistent Homology to Graphs}
\label{subsec:relatedWork}
Huang~\citeyearpar{linkPredictionGraphTopology} discussed the use of topological features such as clustering coefficients link prediction in social networks. Carsten et al.~\citeyearpar{carstens2013persistent} used persistent homology of clique complex of social networks to distinguish them from other network type. Topological features such as simplicial closure have been applied to study community formation in collaboration networks~\citep{patania2017shape}. Persistent homology based methods have recently been used to study temporal evolution of networks and understand the differences in network structure with time~\citep{pal2017comparative,hajij2018visual}.
Benson et al.~\citeyearpar{benson2018simplicial} used simplicial closure of clique complex of graphs to predict link between set of nodes. 
In contrast to all the above works, our method applies persistent homology to understand how an adaptive-sized extended neighborhood of query nodes differ with respect to their homology class. Our method is simple yet robust and achieves performance comparable to sate-of-the-art methods.

\section{Persistence Diagram of a Graph}
\label{sec:pd}
We have presented a general introduction to Persistent Homology in the \ref{ap:ph}. Consider a graph $G=(V,E)$, where $V = \{v_i\}^{n}_{i=1}$ is the node set and $E = \{e_i\}^{m}_{i=1}$ is the edge set. We associate a positive weight $w_{e_i} \in \mathbb{R}, w_{e_i} > 0$ with each of the elements $e_i \in E$. For an \textit{unweighted} graph, $w_{e_i} = 1, \forall {e_i} \in E$. If two nodes are not connected by an edge, we take the (virtual) edge-weight between them as $\infty$, which for practical purposes is taken as a large positive real number $M \in \mathbb{R}$. The \textit{\spd} $D_{sp}(v_i,v_j)$ between the nodes $v_i,v_j \in V$ is defined as the sum of weights of the edges on the path starting at $v_i$ and terminating at $v_j$.

Now consider the metric space $(\mathcal{X},d)$ equipped with a metric $d$. Let $X = \{x_i\}^{n}_{i=1}$ be a set of points in $(\mathcal{X},d)$ such that the points in $X$ correspond to the nodes in $V = \{v_i\}^{n}_{i=1}$. In an \textit{undirected} graph, where the \spd $D_{sp}$ between any two nodes is symmetric, it makes a natural choice for a metric. We can verify that $D_{sp}$ satisfies all the properties of a metric: for arbitrary $v_i,v_j,v_k \in V$, (a) $D_{sp}(v_i,v_j) \geq 0$, (b) $D_{sp}(v_i,v_j) = 0$ $\iff$ $v_i=v_j$, (c) $D_{sp}(v_i,v_j) = D_{sp}(v_j,v_i)$ and (d) $D_{sp}(v_i,v_j) + D_{sp}(v_j,v_k) \geq D_{sp}(v_i,v_k)$. Therefore, for points $x_i,x_j \in X$, which correspond to $v_i,v_j \in V$, we take the metric as $d(x_i,x_j) = D_{sp}(v_i,v_j)$.

For a \textit{directed} graph, the \spd between two nodes is not symmetric. In this case, $d(x_i,x_j) = D_{sp}(v_i,v_j)$  provides a \textit{quasi-metric}: it satisfies (a), (b) and (d) as described above. From a quasi-metric $d(x_i,x_j)$, we derive a metric as follows: $$f_a(x_i,x_j) = a{\times}d(x_i,x_j) + (1-a){\times}d(x_j,x_i)$$ where $a \in [0,1/2]$ \citep{turner2016generalizations}. For $a=\frac{1}{2}$, $f_a(x_i,x_j)$ is the average of the two directed distances. In this work, for a metric space representation of a directed graph, we take $d(x_j,x_i) = \frac{D_{sp}(v_i,v_j) + D_{sp}(v_j,v_i)}{2}$, where $x_i,x_j \in X$ correspond to $v_i,v_j \in V$.% This representation also provides a necessary stability to our algorithm as we explain in the \ref{subsec:ripsfiltration}.

Computing the all-pair-shortest-path in an undirected graph \citep{johnson1977efficient} gives a symmetric \textit{distance matrix} $D = \{d_{ij}\}^{n,n}_{i=1,j=1}$. For a directed graph, the distance matrix is not symmetric; therefore, to impose a metric structure we apply the aforementioned method: $d_{ij} = d_{ji} = \frac{d_{ij} + d_{ji}}{2}$. With that, we have a complete pipeline to compare the shape-features of two graphs (or subgraphs) via their persistent homology.

\section{Link Prediction via Persistent Homology}
\label{sec:ouralgo}
Having discussed the background to compute the quantitative differences between a pair of subgraphs with respect to their shape-features, we describe how to use that to understand and predict the existence of a potential link.
\begin{algorithm}[!htp]
		\footnotesize
%\begin{subfigure}{.5\textwidth}	
	\hspace*{\algorithmicindent} \textbf{Input:} Graph $G$, persistence-threshold $\tau$, a boolean $isD$ to indicate if the graph is directed.% \\
	%\hspace*{\algorithmicindent} \textbf{Output:} 0-dimensional Persistence Diagram of $G$.
\begin{algorithmic}[1]
\renewcommand{\algorithmicprocedure}{\textbf{Algorithm}}	
	\Procedure{\getpd}{$G,\tau,isD$}\label{getpdstart}
\State{$D \gets \getdm(G)$;}\label{algo:getpd:getdm}\Comment{$D = \{D_{sp}(v_i,v_j)\}^{i=|V|,j=|V|}_{i=1,j=1}$. $D_{sp}(v_i,v_j) = M$, if node $v_j$ is unreachable from the node $v_i$, where $M$ is a finite large value.}	
\If{$isD$}
\For{$1 \leq i \leq |V|$, $1 \leq j \leq |V|$}
%\For{$1 \leq j \leq |V|$}
	\State{$d_{ij} \gets d_{ji} \gets \frac{d_{ij} + d_{ji}}{2}$;}\label{algo:getpd:avg}
%\EndFor
\EndFor
\EndIf
\State{$pd \gets \getpddm(D,\tau)$;~Output $pd$;}\label{algo:getpd:pd}
\EndProcedure\label{getpdend}
%	\algstore{addv}
\end{algorithmic}\caption{Computing Persistence Diagram}\label{algo:getpd}
\end{algorithm}

Algorithm \ref{algo:getpd} summarizes the entire pipeline of computing a persistence diagram for a graph $G$. First, we compute the all-pair-shortest-path distance matrix $D$ using any optimized algorithm \citep{johnson1977efficient} for the purpose, line \ref{algo:getpd:getdm}. In case $G$ is directed, $D$ is made symmetric, line \ref{algo:getpd:avg}. Thereafter, $D$ and the persistence-threshold $\tau$ are used to compute the \pd of $G$, line \ref{algo:getpd:pd}, see \citep{bauer2016ripser} for an algorithm to compute a \pd.
\begin{figure}[!htp]	
	\small
	\centering\def\svgwidth{0.9\columnwidth}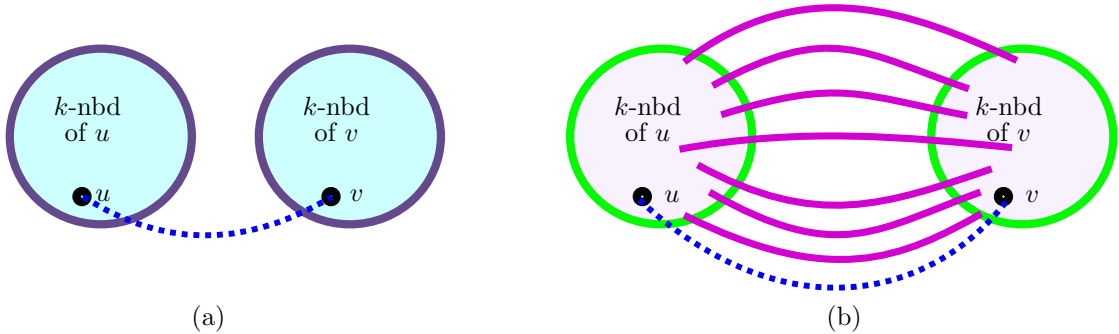
	\caption{\Cnbd of $u$ and $v$, when they have (a) no edges connecting (b) multiple edges connecting.}
	\label{fig:nbd_cases}
\end{figure}

Now consider the cases of \cnbd of nodes $u$ and $v$ as shown in the \ref{fig:nbd_cases}. We consider two scenarios with respect to  reasonably extended neighborhoods of the two nodes, as shown in the \ref{fig:nbd_cases} (a) and (b). 

Essentially, a case of predicting a link between an arbitrary pair of nodes lies on the spectrum of scenarios starting at the one shown in the \ref{fig:nbd_cases} (a) and stretches towards the ones similar to the \ref{fig:nbd_cases} (b). As we mentioned in \ref{subsec:insight}, the existence of a possible link has higher chances as we move away from the case of the \ref{fig:nbd_cases} (a) on this spectrum.

With that observation, we explore and understand how the difference in shape-measures, as provided by the distances in the \pds of a number of subgraphs induced by the \cnbd of $u$ and $v$, varies when we examine the cases of arbitrary pair of nodes. This is presented in the \ref{alg:getdis}.
\begin{algorithm}[!htp]
	\footnotesize
	\hspace*{\algorithmicindent} \textbf{Input:} Graph $G$, Nodes $u,v$, Neighborhood radius $r$, Combined-neighborhood radius $k$, persistence-threshold $\tau$, a boolean $isD$ to indicate if directed.% \\
	%\hspace*{\algorithmicindent} \textbf{Output:} Nodes in the $r-$neighborhood of $v$ sorted by the proposed similarity measure.
	\begin{algorithmic}[1]
		\renewcommand{\algorithmicprocedure}{\textbf{Algorithm}}	
		\Procedure{\getdis}{$G,u,v,k,\tau, isD$}\label{getdisstart}

\State{$N^k_{u} \gets \getnbd(u,k)$;}\Comment{Induced subgraph over $i$-hop neighbors of $n$, where $1{\leq}i{\leq}k$.}\label{alg:getdis:nu}
\State{$N^k_{v} \gets \getnbd(v,k)$;}\label{alg:getdis:nv}%\Comment{Induced subgraph over $i$-hop neighbors of $n$, where $1{\leq}i{\leq}k$.}	
\State{$N^{k}_{u,v} \gets \getcnbd(u,v,j)$;}\label{alg:getdis:nuv}\Comment{Induced subgraph over $i$-neighbors of $v$ or $n$ or both, where $1{\leq}i{\leq}k$.}	
\State{$N^{k+}_{u,v} \gets N^{k}_{u,v} \cup (u,v)$;}\label{alg:getdis:nupv}\Comment{Induced subgraph $N^{k}_{u,v}$ augmented with the edge $(u,v)$.}	
\State{$N^{k-}_{u,v} \gets N^{k}_{u,v} - (u,v)$;}\label{alg:getdis:numv}\Comment{Induced subgraph $N^{k}_{u,v}$ without the edge $(u,v)$.}	
\State{$C(N^{k}_{u,v}) \gets \mkcom(N^{k}_{u,v})$;}\label{alg:getdis:nc}\Comment{The complete graph over the nodes of $N^{k}_{u,v}$.}	
\State{$P_u \gets \getpd(N^k_{u},\tau,isD)$;}\Comment{Persistence diagram of the subgraph induced by $N^{k}_u$.}	
\State{$P_v \gets \getpd(N^k_{v},\tau,isD)$;~$P^+ \gets \getpd(N^{k+}_{u,v},\tau,isD)$;}%\Comment{Persistence diagram with threshold $\tau$ of the subgraph $N^{k+}(u,v)$.}	
\State{$P^- \gets \getpd(N^{k-}_{u,v},\tau,isD)$;~$P^c \gets \getpd(C(N^{k}_{u,v}),\tau,isD)$;}%\Comment{Persistence diagram with threshold $\tau$ of the subgraph $C(N^{k}_{u,v})$.}	
%\State{$\{d_1,d_1,d_1,d_1\} \gets \{\getwd(P^+, P^-),\getwd(P^+, P^c)\linebreak,\getwd(P^+, P^c),\getwd(P^+, P^c)\}$;}
\State{$d_1 \gets \getwd(P^+, P^-)$;~$d_2 \gets \getwd(P^+, P^c)$;}\label{alg:getdis:do}
\State{$d_3 \gets \getwd(P^+, P_u)$;~$d_4 \gets \getwd(P^+, P_v)$;}
\\\Comment{Wasserstein$-2$ distances between the Ps}		
\State{$d_5 \gets \getbd(P^+, P^-)$;~$d_6 \gets \getbd(P^+, P^c)$;}
\State{$d_7 \gets \getbd(P^+, P_u)$;~$d_8 \gets \getbd(P^+, P_v)$;}\label{alg:getdis:dt}
\\\Comment{Bottleneck distances between the Ps}		
\State{$\tilde{d} \gets \{d_1,d_2,d_3,d_4,d_5,d_6,d_7,d_8\}$;}\Comment{A vector of the eight distances.}	
\State{Output $\tilde{d}$;}
\EndProcedure\label{getdisend}
\end{algorithmic}\caption{Getting vector of Bottleneck and Wasserstein-$2$ distances.}\label{algo:getnodes}
\label{alg:getdis}
\end{algorithm}

Given a graph $G = (\{v_i\}^{n}_{i=1},\{e_k\}^{m}_{k=1})$, for a $k \leq n$, first, we compute the subgraph of $G$ induced by the $i$-hop neighbors of $u$ and $v$, where $1 \leq i \leq k$, see lines \ref{alg:getdis:nu} and \ref{alg:getdis:nv}. Thereafter, we compute the subgraph induced by the union of the two neighborhoods, see line \ref{alg:getdis:nuv}. From this subgraph, we induce two subgraphs corresponding to the existence and non-existence of a link between the query nodes, see  lines \ref{alg:getdis:nupv} and \ref{alg:getdis:numv}. Following our intuition, a missing link in a complete graph has high chances of existence, therefore, we also compute a complete graph over the nodes of the combined \nbd, line  \ref{alg:getdis:nc}. Having collected these subgraphs, we compute their \pds as detailed in \ref{algo:getpd}.

In the \pds, we have considered only $0^{th}$ persistent homology group. This is because the cycles in a graph, which correspond to its $1^{st}$ persistent homology group, are never destroyed as there are no 2-faces. Thus, for our purpose distances between the 1-dimensional \pds of two subgraphs would not help much.

We compute the \wst and \bn distances between the \pds, as shown in the lines \ref{alg:getdis:do} to \ref{alg:getdis:dt}. They signify how much the induced \textit{subgraphs are similar} with respect to their shape-features, in case (a) the target link exists: inferred by $d_1$ and $d_5$ (b) the \nbd of one of the query nodes contains the \nbd of the other one: inferred by $d_3$, $d_4$, $d_7$ and $d_8$, and (c) the \cnbd of the two query nodes is very close to a complete graph: inferred by $d_2$ and $d_6$. These different views of contrasting the subgraphs enable us to understand and thereby predict the existence of a link between any arbitrary pair of nodes. We use the vector of these eight distance measures in our experiments as described in the next section.

\section{Experiments}
\label{sec:exp}
\subsection{Datasets}
We use following publicly available network datasets to evaluate the strengths and weaknesses of the proposed method for the link prediction task.

\noindent
\textbf{David Copperfield (DC):}\footnote{http://www-personal.umich.edu/~mejn/netdata/adjnoun.zip}~\citep{dc} This is an adjacency network of nouns and adjectives extracted from Charles Dickens' novel \textit{David Copperfield}.

\noindent
\textbf{Air Traffic Control (ATC):}\footnote{http://research.mssm.edu/maayan/datasets/qualitative\_networks.shtml} This network is constructed from the Preferred Route Database (PRD) as provided by the National Flight Data Center of the US Federal Aviation Agency. Nodes represent an airport or a service centre and edges correspond to the routes as present in PRD.

\noindent
\textbf{CORA}\footnote{https://linqs-data.soe.ucsc.edu/public/lbc/cora.tgz}~\citep{cora}: It is a citation network of 2708 papers (nodes) and 5492 citations (edges).

\noindent
\textbf{Yeast}\footnote{http://moreno.ss.uci.edu/data.html\#pro-pro}\citep{moreno} represents protein-protein interactions as found in Yeast.

\noindent
\textbf{Power}\footnote{http://konect.uni-koblenz.de/networks/opsahl-powergrid}~\citep{power}: This network represents the power grid of Western United States with each node corresponding to sub-station and edges representing the power supply lines connecting the sub-stations.  

\noindent
\textbf{Gnutella Peer to Peer Network (p2p)}\footnote{https://snap.stanford.edu/data/p2p-Gnutella04.html}~\citep{p2p} represents a snapshot of the Gnutella file sharing peer to peer network where each node corresponds to a host in the network and edges correspond to the connections between hosts.

\noindent
\textbf{Twitter Follower Network (Twitter):}\footnote{http://snap.stanford.edu/data/egonets-Twitter.html}~\citep{twitter} In this network, each node corresponds to a Twitter user and and an edge from user $A$ to user $B$ indicates that $B$ is followed by $A$.

\begin{table}[ht]
	\centering
	%\resizebox{0.8\columnwidth}{!}{%
		\begin{tabular}{@{}lccl@{}}
			\toprule
			& \textbf{\# nodes} & \textbf{\# edges} & \textbf{N/w Type}\\ \midrule
			\textbf{DC} & 112 & 425 & Word Co-occurrence n/w \\
			\textbf{ATC} & 1226 & 2615 & Air Traffic n/w \\
			\textbf{Cora} & 2708 & 5429 & Citation n/w \\
			\textbf{Yeast} & 1870 & 2277 & Protein-protein interaction n/w \\
			\textbf{Power} & 4941 & 6594 &  Power Grid n/w\\
			\textbf{p2p} & 10876 & 39994 & Peer to peer network \\
			\textbf{Twitter} & 23370 & 33101 & Social Network\\ \bottomrule
		\end{tabular}%
	%}
	\caption{Different datasets used in experiments}
	\label{dataStats}
\end{table}
\subsection{Baselines}
\noindent
\textbf{Adamic-Adar (AA)}~\citep{adamicAdar} is one of the most commonly used method for link prediction, especially in social networks. This method is based on the intuition that two persons should be friends (linked) if they share many friends in common.

\noindent
\textbf{Mile-Witten (MW)}~\citep{milneWitten} Is another frequently used link prediction heuristic that was proposed in context of measuring semantic relatedness of Wikipedia articles and is based on the intuition that two Wikipedia articles are topically related
if there are many Wikipedia articles that link to both of them.

\noindent
\textbf{DeepWalk (DW)}~\citep{deepwalk} learns latent representations of nodes in the graph by utilizing local information obtained from truncated random walks.

\noindent
\textbf{Node2Vec}~\citep{node2vec} maps nodes in a network in a low dimensional feature space that preserves the neighborhood structure of the nodes. Different from DeepWalk, it uses biased random walks to explore diverse neighborhoods.

\subsection{Experimental Protocol}
\label{sec:protocol}
We implemented our proposed algorithm in C++ and used the Ripser library\footnote{{https://github.com/Ripser/ripser}} for computing persistence diagrams. The methods for computing the \bn and \ws distances are taken from the publicly available code\footnote{https://bitbucket.org/grey\_narn/hera/src/master/} of \citep{kerber2016geometry}. We also implemented the baselines AA and MW in C++ and used author provided source code for node2vec\footnote{https://github.com/aditya-grover/node2vec} and DeepWalk\footnote{http://www.perozzi.net/projects/deepwalk/}.
For all the datasets, we created test-train splits by holding out roughly 5\% of edges and used 95\% of the edges for training. 

\subsection{Link Prediction as a Ranking Task}
\begin{table}[t]
\centering
\resizebox{\columnwidth}{!}{
\begin{tabular}{@{}cllllllll@{}}
\toprule
 &  & \textbf{DC} & \textbf{ATC} & \textbf{Cora} & \textbf{Power} & \textbf{Yeast} & \textbf{p2p} & \textbf{Twitter} \\ \midrule
\multirow{5}{*}{\textbf{Hits @ 1}} & AA & \textbf{0.095} & 0.000 & \textbf{0.040} & \textbf{0.039} & 0.017 & 0.000 & 0.001 \\
 & MW & 0.047 & \textbf{0.015} & 0.014 & 0.012 & 0.012 & 0.000 & 0.004 \\
 & DeepWalk & 0.000 & \textbf{0.015} & 0.032 & 0.015 & \textbf{0.055} & 0.000 & 0.000 \\
 & node2vec & 0.000 & 0.007 & 0.000 & 0.000 & 0.027 & 0.000 & 0.000 \\
 & Topology & 0.000 & 0.000 & 0.030 & 0.021 & 0.026 & 0.000 & \textbf{0.008} \\
 \midrule
\multirow{5}{*}{\textbf{Hits @ 10}} & AA & 0.142 & 0.046 & 0.072 & 0.200 & \textbf{0.132} & 0.000 & 0.047 \\
 & MW & 0.095 & 0.030 & 0.034 & 0.161 & \textbf{0.132} & 0.000 & 0.045 \\
 & DeepWalk & \textbf{0.238} & \textbf{0.200} & \textbf{0.362} & 0.221 & 0.083 & \textbf{0.004} & 0.012 \\
 & node2vec & 0.095 & 0.061 & 0.022 & 0.000 & 0.027 & 0.000 & 0.000 \\
 & Topology & 0.190 & 0.023 & 0.272 & \textbf{0.231} & \textbf{0.132} & 0.002 & \textbf{0.048} \\
 \midrule
\multirow{5}{*}{\textbf{Hits @ 50}} & AA & 0.238 & 0.061 & 0.072 & 0.237 & 0.238 & 0.000 & 0.053 \\
 & MW & 0.095 & 0.069 & 0.034 & 0.227 & 0.283 & 0.000 & 0.158 \\
 & DeepWalk & 0.523 & \textbf{0.446} & \textbf{0.674} & \textbf{0.534} & \textbf{0.388} & \textbf{0.069} & \textbf{0.188} \\
 & node2vec & 0.38 & 0.169 & 0.082 & 0.015 & 0.027 & 0.002 & 0.001 \\
 & Topology & \textbf{0.571} & 0.100 & 0.328 & 0.504 & 0.345 & 0.007 & 0.131 \\ 
\bottomrule 
\end{tabular}%
}\caption{Performance of different methods on seven different datasets when treating link prediction as a ranking task. Hits at ranks 1,10, and 50 are reported. For each dataset, the best method achieving highest hits at a given rank is highlighted in bold.}
\label{tab:rankingResults}
\end{table}
Traditionally, the problem of link prediction has been addressed as a ranking problem where given a source node, a ranked list of target nodes is produced ordered by the likelihood of a potential link being formed between the source and the target node~\citep{kleinbergLinkPrediction,linkPredRank2,linkPredRank}.  The baselines AA and MW, by definition, output a score between the source and target node that can be used as the ranking function. The other two baselines -- deepwalk and node2vec -- learn continuous vector representations for each node in the graph. A typical way to rank target nodes given a source node is to rank them based on their distance from the source node~\citep{misraAAAI18}. Hence, for these two methods, given a source node, we produce a ranked list of all the other nodes in the graph ordered by the Euclidean distance between the source and target node vectors.

Given a pair of source and target nodes our proposed approach produces eight different distance values (Algorithm~\ref{alg:getdis}) capturing the structural similarities between the extended neighborhood of the source and target nodes. While each distance value can independently be used to rank the target nodes, we are interested in producing a ranked list that combines the different properties captured by the different distance functions. In order to do that, we use the \emph{rank product} metric~\citep{rankProduct} to combine the ranked lists produced by individual distance functions to obtain the final ranking of target nodes with respect to a given source node. For a node $i$, the rank product is computed as
\begin{equation}
rp_i = (\prod_{j=1}^m r_{ij})^{1/m}
\end{equation}

where $r_{ij}$ is the rank of node $i$ in the $j^{th}$ ranked list. Table~\ref{tab:rankingResults} reports the results achieved by the four baseline methods and the ranking produced by our proposed approach. We use \emph{Hit Rate}$ @N$, i.e., the proportion of edges for which the correct target node was ranked in the top $N$ positions. We report hit rate for $n=\{1,10,50\}$.  It can be observed from the table that no single method achieves consistent outperforms across all metrics for the different datasets. This is consistent with previous empirical studies comparing performance of different link prediction techniques~\citep{kleinbergLinkPrediction,complexNetSurvey}.  In terms of Hit Rate at rank 10, rankings produced by deepwalk achieve the best performance for four out of seven datasets and rankings produced by the proposed approach achieve the best performance for the remaining three datasets. Further, we note that the performance of rankings produced by our proposed approach and the deepwalk embeddings are relatively more stable and robust compared to the other baselines across different datasets.  This strong performance achieved by the proposed approach is commendable given that the proposed approach uses only eight features (different distance functions capturing the topological properties) that can be computed with relative ease compared to computationally expensive learning of vector representations (as is the case with deepwalk and node2vec). Further, unlike the AA and MW baselines, which are also easier to compute, the proposed approach is built upon the solid theoretical foundations of topological data analysis and persistent homology.  

\section{Conclusions and Future Work}
\label{sec:discuss}
We proposed an approach inspired from persistent homology to model link formation in graphs and use it to predict missing links. The proposed approach achieved robust and stable performance across seven different datasets. The performance achieved was comparable to the state-of-the-art baselines despite the proposed approach being relatively simple and computationally less expensive than some of the baselines that also utilize global network information. Given that the topological features succinctly capture information about shape and structure of the network and can be computed without the need of extensive training, it will be worth exploring how these features can be combined with other techniques for network analysis.

\section*{Acknowledgment} We thank Dr. Yogeshwaran D., Indian Statistical Institute Bangalore, India, for valuable discussions on persistent homology of graphs. 

\bibliographystyle{named}
\bibliography{references}
\appendix
\section{Persistent Homology}
\label{ap:ph}
For a self-contained exposition, here we present the definitions of the main notions used in this work. We assume a familiarity with the elementary Group Theory e.g. definition of group, abelian group, quotient group, group homomorphism, kernel, rank, etc.~\citep{gallian2012contemporary}. An interested reader can refer to~\citep{zomorodian2005computing,edelsbrunner2008persistent,edelsbrunner2010computational} for further reading on persistent homology.

\textbf{Abstract simplicial complex:} Given a set $S$, consider a family of subsets $\Delta \subseteq 2^S$. $\Delta$ is called an \textit{abstract simplicial complex} if $\forall \rho \in \Delta$, $\kappa \subset \rho$ $\implies$ $\kappa \in \Delta$. Each of the sets $\rho \in \Delta$ of size $|\rho| = (n+1)$ is called an \textit{$n$-simplex}. A subset $\varepsilon \subsetneq \rho$ is called a \textit{face} of $\rho$.	

\textbf{Simplicial Homology:} Let $\Delta$ be a simplicial complex. A \textit{simplicial $n$-chain}, where  $n \in \mathbb{Z}$, is a sum of $n$-simplices of $\Delta$ as given below: $$c = \sum^{k}_{i=1}\alpha_i\rho_i, \alpha_i \in \mathbb{Z}, \rho_i \in \Delta, |\rho_i| = (n+1)$$ 
The set of $n$-chains denoted as $C_n(\Delta)$ or just $C_n$ is an \textit{abelian group} with the sum as defined above, identity element as the \textit{null chain} i.e. $c$ where each of the $\alpha_i$ are 0, and the inverse of each element as the element itself.

The \textit{boundary} of an $n$-simplex $\rho=\{s_o,s_1,\ldots,s_n\}$ is defined as the chain of its $(n-1)$-faces: $\delta_n(\rho) = \sum_{i=0}^{n}\{s_o,s_1,\ldots,s_{i-1},s_{i+1},\ldots,s_n\}$. 
The \textit{$n$-boundary} of a simplicial complex is defined as the sum of the boundaries of its $n$-simplices. We denote the set of $n$-boundaries of $\Delta$ as $B_n(\Delta)$, or just $B_n$ in the context of the same $\Delta$.

For $n$-chains $c_i,c_j \in C_n$, $h_n: C_n \rightarrow C_{n-1}$ is a \textit{group homomorphism}: $c_i,c_j \in C_n \Rightarrow h_n(c_i+c_j) = h_n(c_i) + h_n(c_j)$. With this definition of $h_n$, we define a \textit{chain-complex} as the following sequence of abelian groups:
$$\ldots \xrightarrow{h_{n+2}}C_{n+1} \xrightarrow{h_{n+1}}C_{n} \xrightarrow{h_{n}}\ldots \xrightarrow{h_{2}}C_{1} \xrightarrow{h_{1}}C_{0} \xrightarrow{h_{0}}0$$
A $c \in C_n$ is called an \textit{$n$-cycle} if $h_n(c) = 0$. The set of all the $n$-cycles of $\Delta$ is denoted by $Z_n(\Delta)$. It can be observed that $Z_n(\Delta) \subseteq C_n(\Delta)$ and $Z_n(\Delta)$ is a \textit{kernel} of $h_n$. Thus $Z_n(\Delta)$ is a subgroup of $C_n(\Delta)$.	It can also be observed that the set of $n$-boundaries $B_n$ is a subgroup of $C_n$. 

Please note that $B_n$ is the image of the group homomorphism $h_{n+1}$. Furthermore, for any $n \in \mathbb{Z}$, $h_{n-1} \circ h_n = 0$. It follows that $B_n(\Delta) \subset Z_n(\Delta)$. Consequently, we (validly) define the $n^{th}$ \textit{simplicial homology group} of $\Delta$ as the quotient group $H_n(\Delta) = Z_n(\Delta)/B_n(\Delta)$.

The $n^{th}$ \textit{Betti number} $\beta^n$ of a simplicial complex $\Delta$ is defined as the \textit{dimension} of its $n^{th}$ {simplicial homology group} $H_n(\Delta)$. Therefore, $\beta^n = rank(H_n(\Delta)) = rank(Z_n(\Delta)) - rank(B_n(\Delta))$.
An element of $H_n(\Delta)$ is called an \textit{$n^{th}$-dimensional homology class} of $\Delta$. Thus $\beta^n$ counts the number of {$n^{th}$-dimensional homology class}.% of $\Delta$.

\textbf{Filtration:} Let $\Delta$ be a finite abstract simplicial complex. A finite \textit{nested} sequence $\{\Gamma_i\}_{i \in I}$ of \textit{sub-complexes} of $\Delta$ is called its \textit{filtration}. To construct a filtration of $\Delta$, we take a real-valued function $f : \Delta \rightarrow \mathbb{R}$ such that for $\rho,\sigma \in \Delta$, $f(\rho) \leq f(\sigma)$ whenever $\rho \subset \sigma$. With that, a \textit{sublevel set} $\Gamma_r = f^{-1}((-\infty,r])$ for $r \in \mathbb{R}$ is a sub-complex of $\Delta$. By the definition of $f$, ordering the sub-complexes by the increasing values of $r$ gives a filtration of $\Delta$ as the following: $$\Phi = \Gamma_0 \subsetneq \Gamma_1 \subsetneq \Gamma_2 \ldots \subsetneq \Gamma_p = \Delta$$ 

\textbf{Persistence Diagram:} For a pair of $i,j~s.t.~ 0 \leq i \leq j \leq p$, the above inclusion relation among $\Gamma_i$s induces a homomorphism on the simplicial homology group of each dimension $n \in \mathbb{Z}$ given by $f^{i,j}_n : H_n(\Gamma_i) \rightarrow H_n(\Gamma_j)$.

The $n^{th}$ \textit{persistent homology group} is the image of the homomorphism $f^{i,j}_n$ given by $Im(f^{i,j}_n)$. In turn, the $n^{th}$ \textit{persistent Betti number} is defines as the rank of $Im(f^{i,j}_n)$ given by $\beta^{i,j}_n = 	rank(Im(f^{i,j}_n))$.

The $n^{th}$ {persistent Betti number} counts how many homology classes of dimension $n$ survives a passage from $\Gamma_i$ to $\Gamma_j$. We say that a homology class $\alpha \in  H_n(\Gamma_i)$ is \textit{born} at resolution $i$ if it did not come from a previous sub-complex: $\alpha \notin Im(f^{i-1,i}_n)$. Similarly, we say that a homology class dies at resolution $j$ if it does not belong to the sub-complex $\Gamma_j$ and belonged to previous sub-complexes.

\begin{wrapfigure}{r}{2cm}
	\footnotesize
	\centering
	\def\svgwidth{2cm}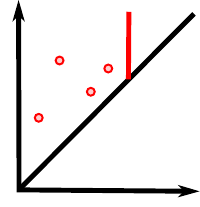
	\caption{PD}
	\label{fig:pd}
\end{wrapfigure}
A \textit{persistence diagram} is a plotting of the points $(i,j)$ corresponding to the birth and death resolutions, respectively, for each of the homology classes. Because a homology class can not die before it is born, all the points $(i,j)$ lie above the diagonal $x=y$. If a homology class does not die ever after its birth, we draw a vertical line starting from the diagonal in correspondence to its birth. For practical purposes, we take a \textit{persistence threshold} $\tau$, and assume that every homology class dies at the resolution $\tau$. A typical persistence diagram is shown in the \ref{fig:pd}.

\textbf{Distance between Persistence Diagrams:} A \pd can be considered a summary of \textit{topological features} associated with the homology groups of the simplicial complex. So, to compare the topological features, we quantify the difference between the \pds in terms of the distance measures.

Let $P_1$ and $P_2$ be two \pds. Let $\eta$ be a bijection between the points in the two diagrams. We define the following two distance measures:
\begin{enumerate}[label=(\alph*)]
	\item \textit{Wasserstein-$q$ distance}: \begin{small}$$W_q(P_1,P_2) = \left[\inf_{\eta:P_1{\rightarrow}P_2}\sum_{p{\in}P_1}||p-\eta(p)||_\infty^q\right]^{\frac{1}{q}}$$\end{small}
	\item \textit{Bottleneck Distance}: \begin{small}$$W_\infty(P_1,P_2) = \inf_{\eta:P_1{\rightarrow}P_2}\sup_{p{\in}P_1}||p-\eta(p)||_\infty$$\end{small}
\end{enumerate}
The \ws distance is sensitive to small differences in the \pds, whereas, the \bn distance captures relatively large differences.

\begin{figure}[!htp]	
	\small
	\centering\def\svgwidth{0.9\columnwidth}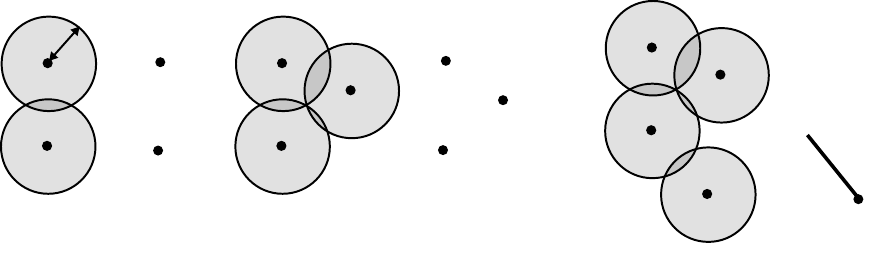
	\caption{Rips complexes containing: (a) a 1-simplex (b) a 2-simplex (c) a 1-simplex and a 2-simplex.}
	\label{fig:rips}
\end{figure}

\textbf{Rips Complex:} A Vietoris-Rips Complex, also called a \textit{Rips complex} is an abstract simplicial complex defined over a finite set of points $X = \{x_i\}^{n}_{i=1} \subseteq \mathcal{X}$ in a metric space $(\mathcal{X},d)$. Given $X$ and a real number $r > 0, r \in \mathbb{R}$, a Rips complex $R(X,r)$ is formed by connecting the points for which the balls of radius \rbt centered at them intersect. In the context of the same point set, we use $R_r$ to denote $R(X,r)$. A 1-simplex is formed by connecting two such points and corresponds to an edge. A 2-simplex is formed by 3 such points and corresponds to a triangular face. See \ref{fig:rips}

\textbf{Rips Filtration:} Given a set of points $X = \{x_i\}^{n}_{i=1} \subseteq \mathcal{X}$, let $0=r_0 \leq r_1 \leq r_2 \ldots \leq r_m$ denote a finite sequence of increasing real numbers, which we use to construct Rips complexes $\{R_{r_i}\}^m_{i=1}$ as defined above. Clearly, by construction of Rips complexes the sequence $\{R_{r_i}\}^m_{i=1}$ is nested and thus provides a filtration of $R_{r_m}$:
$$\Phi = R_{r_0} \subsetneq R_{r_1} \subsetneq R_{r_2} \ldots \subsetneq R_{r_m}$$	

Deriving the persistent homology groups via homomorphism over a Rips filtration, we obtain a persistence diagram associated with the point set $X$. Please note that to compute the Rips filtration associated with a point set $X$ we need only the relative pairwise distances between the points $x_i \in X$. Essentially, we need a symmetric distance matrix $D = \{d(x_i,x_j)\}^{n,n}_{i=1,j=1}$ to compute the persistence diagram of $X$. Next, we will use this method to compute the persistence diagram of a graph.

\end{document}